\pdfoutput=1
\documentclass[aps,pre, twocolumn, groupedaddress]{revtex4-1}
\usepackage{lipsum}
\usepackage{mathtools}
\usepackage{graphicx}
\usepackage{dcolumn}
\usepackage{amsmath}    
\usepackage{amssymb}
\usepackage{bm}
\usepackage{hyperref}
\usepackage{latexsym}
\usepackage{verbatim}
\usepackage{color}
\usepackage{float}
\usepackage[caption=false]{subfig}
\usepackage{epstopdf}

\def\beq{\begin{equation}}
\def\eeq{\end{equation}}

\def\beq{\begin{equation}}                          
\def\eeq{\end{equation}}                          
\def\bea{\begin{eqnarray}}                          
\def\eea{\end{eqnarray}}

\DeclareRobustCommand{\uvec}[1]{{%
  \ifcsname uvec#1\endcsname
     \csname uvec#1\endcsname
   \else
    \bm{\hat{\mathbf{#1}}}%
   \fi
}}

\preprint{}

\begin{document}

\title{Dynamics of particle moving in a two dimensional Lorentz lattice gas}

\author{Pranay Bimal Sampat}
\email[]{pranayb.sampat.phy16@itbhu.ac.in}
\affiliation{Department of Physics, Indian Institute of Technology (BHU), Varanasi, U.P. India - 221005}

\author{Sameer Kumar}
\email[]{sameerk.rs.phy16@itbhu.ac.in}
\affiliation{Department of Physics, Indian Institute of Technology (BHU), Varanasi, U.P. India - 221005}

\author{Shradha Mishra}
\email[]{smishra.phy@itbhu.ac.in}
\affiliation{Department of Physics, Indian Institute of Technology (BHU), Varanasi, U.P. India - 221005}

\begin{abstract}
	We study the dynamics of a  particle moving in a square two-dimensional Lorentz lattice-gas. The underlying lattice-gas is occupied by  two kinds of rotators, ``right-rotator (R)'' and ``left-rotator (L)'' and some of the sites are empty {\it{viz.}} vacancy ``V''.The density of $R$  and $L$ are the same and density of $V$ is one of the key parameters of our model. The rotators deterministically rotate the direction of a particle's velocity to the right or left and vacancies leave it unchanged. We characterise the dynamics of particle motion for different densities of vacancies. Since the system is deterministic, the particle forms a closed trajectory asymptotically. The probability of the particle being in a closed or open trajectory at time $t$ is a function of the density of vacancies. \textcolor{black}{The motion of the particle is {\it{uniform}} throughout in a fully occupied lattice. However, it is divided in two distinct phases in partially vacant lattices}: The first phase of the motion, which is the focus of this study, is characterised by anomalous diffusion and a power-law decay of the probability of being in an open trajectory. The second phase of the motion is characterised by subdiffusive motion and an exponential decay of the probability of being in an open trajectory. For lattices with a non-zero density of vacancies, the first phase of motion lasts for a longer period of time as the density of vacancies increases.
\end{abstract}

\maketitle
\section{Introduction}

Most native species, like microscopic particles, encounter random obstacles in the environment in which they move, 
i.e. they move in an {\it inhomogeneous} environment \cite{goldingprl2006, naturemater2013,parry}. 
These obstacles or the environment directly influence the motion of the microscopic particles and can be modeled by 
a pattern of barriers/obstacles in the medium, which control the motion of a particle moving in it. 
Depending on the characteristics of the environment, motion can be of different kinds, e.g. anomalous diffusion or 
confined motion \cite{suryanjop2016, revarticle}. 
 Previous studies have addressed the action of the particle(s) in an 
inhomogeneous environment and its effect on particle dynamics and steady state \cite{stark, subdiffusion}. 
The Lorentz lattice gas (LLG) \cite{205llg, binder1987, binder1988} has turned out to be a useful way to  model different types of 
inhomogeneous environment. Several physical phenomena have been studied using the Lorentz lattice gas 
\cite{langton, gale, troubetskoy, shradhajstatphys, meng1994, benweb, oneDLLG}, in one and two dimensions. \\

In \cite{205llg}, the authors have studied a fixed obstacle model, where the properties of the 
environment remain unchanged throughout the motion of the particle, unlike the 
case of  \cite{shradhajstatphys, oneDLLG} where the environment can change with 
the motion of the particle. For fixed obstacles, it is found that the particle forms a closed trajectory asymptotically\cite{205llg}.\\

In our present study, \textcolor{black}{the obstacles are fixed and are modeled as left and right rotators on a square lattice}, which rotate the direction 
of particle velocity towards the left or right 
respectively. Some of the sites are vacant and the direction of velocity of a particle passing through 
these sites remains unchanged. \textcolor{black}{The configuration of the lattice sites, however, is generated randomly, which adds stochasticity to the model.} The density of vacant sites $\rho_V$ is one of the key parameters and 
the ratio of left/right rotators is kept fixed at one. 
The present model is the same as the one introduced by Cohen {\it et al.} in 1995 \cite{205llg}. In the current study 
we explore the approach to the asymptotic behaviour in detail for many vacancy densities.
For any $\rho_V \ne  1$, the asymptotic behaviour of the particle is a closed trajectory  
as reported in \cite{205llg}. However, we found that the approach to the asymptotic state depends on the density of the vacant sites 
in the lattice. Like the  Ruijgrok-Cohen (RC) mirror model of lattice gas {\cite{rziff}, the presence of vacancies in the rotator model of LLG causes a significant change in the motion of the particle. \\
Here we write them briefly. (i) The motion of the particle in a partially vacant LLG can be separated in two distinct phases of motion. The first phase of the motion of the particle is anomalously diffusive for all values of vacancy density we have considered. The effective diffusion coefficient, however, depends on the vacancy density, and increases monotonically with the density of vacancies. The second phase of the particle's motion in lattices with vacancies is subdiffusive, \textcolor{black}{and the subdiffusion appears at increasingly late times as vacancy density increases.} \\
(ii) The probability of the particle being in an open trajectory decays as a power law with time in the first phase of motion. The magnitude of the power-law exponent decreases on increasing the density of vacancies. For lattices with vacancies, the probability of being in an open trajectory decays exponentially after a period of time, as reported in \cite{205llg}. The motion of the particle after that time is described as the second phase of its motion and \textcolor{black}{the time spent by the particle in the first phase of its motion increases with an increase in the density of vacancies.}
As a result of (i) and (ii), particles in lattices with a higher density of vacancies asymptotically form larger closed trajectories over a longer period of time, compared to a lattice with a lower, non-zero value of vacancy density.  \textcolor{black}{The boundary of the trajectory of the
particle also becomes increasingly rough as vacancy
density increases.}\\

In the rest of the article: in section \ref{model} we discuss the model in detail. \textcolor{black}{In section \ref{det}  we discuss how the deterministic nature of the  lattice leads to periodic trajectories.}
We then discuss the results of our model in section \ref{results}
and finally conclude in section \ref{discussion}.

\section{Model and numerical details} 
\label{model}

We study the dynamics of a {\em single} particle moving along the bonds of a square lattice of unit lattice spacing. The particle covers the distance of the unit bond length in a unit time step.
The lattice  consists of two types of rotators- ``left'' (L) and ``right'' (R), randomly distributed over the lattice with  an equal probability which is independent of the rotators present on the other lattice sites. Some of the sites on the lattice are vacant, and are termed as ``vacancies" (V). 
The density of vacancies is one of the control parameters of our model. L/R rotators change the direction of particle velocity to the left/right (by an angle of $\frac{-\pi}{2}/\frac{\pi}{2}$) respectively, whereas the  velocity 
remains unchanged when it encounters a vacant site. Fig. \ref{fig: 1}(a) shows the cartoon of a part
of the lattice and \ref{fig: 1}(b) shows the  interaction of a particle with L/R rotators and vacancies.
The density of ``L", ``R" and ``V" is defined as $\rho_L$, $\rho_R$ and  $\rho_V$ respectively 
and $\rho_L+\rho_R+\rho_V=1$. For any given value of $\rho_V$, $\rho_L$ and $\rho_R$ are always equal. \\

The particle starts its motion from a point at $t=0$ and the configuration of a lattice site is generated when the particle visits it for the first time, using the uniform probability distribution which is given by densities 
$\rho_L$, $\rho_R$ and $\rho_V$ for left rotators, right rotators and vacancies respectively. 
After this, the configuration of the lattice site is stored and it remains unchanged throughout
the motion, i.e. the motion is deterministic. At each time step, the program checks whether the new position of the particle has been visited earlier. If it has, the velocity of the particle will be updated according to the configuration of the site generated earlier. In the present model, the particle 
never experiences a boundary as the lattice is infinite. 
As time progresses and the number of visited sites increases, the model needs to use more memory to store the information about the visited sites and also takes more time to check whether a site has been visited earlier. However, our approach has the following advantages over saving the whole lattice at once:\\

(i) If we were to generate a finite lattice for which the particle is theoretically always trapped within the boundary, the time required to generate the lattice and the memory required to store it will be of the order of  $O(t_{max}^{2})$, where $t_{max}$ is the total number of time steps.\\

(ii) If we were to generate a smaller finite lattice than the theoretically required lattice, the particle could cross the boundary. In such a case, the code has to be stopped and the realisation has to be excluded when calculating observables.
\begin{figure}
\centering
\includegraphics[width=8cm,height=4cm]{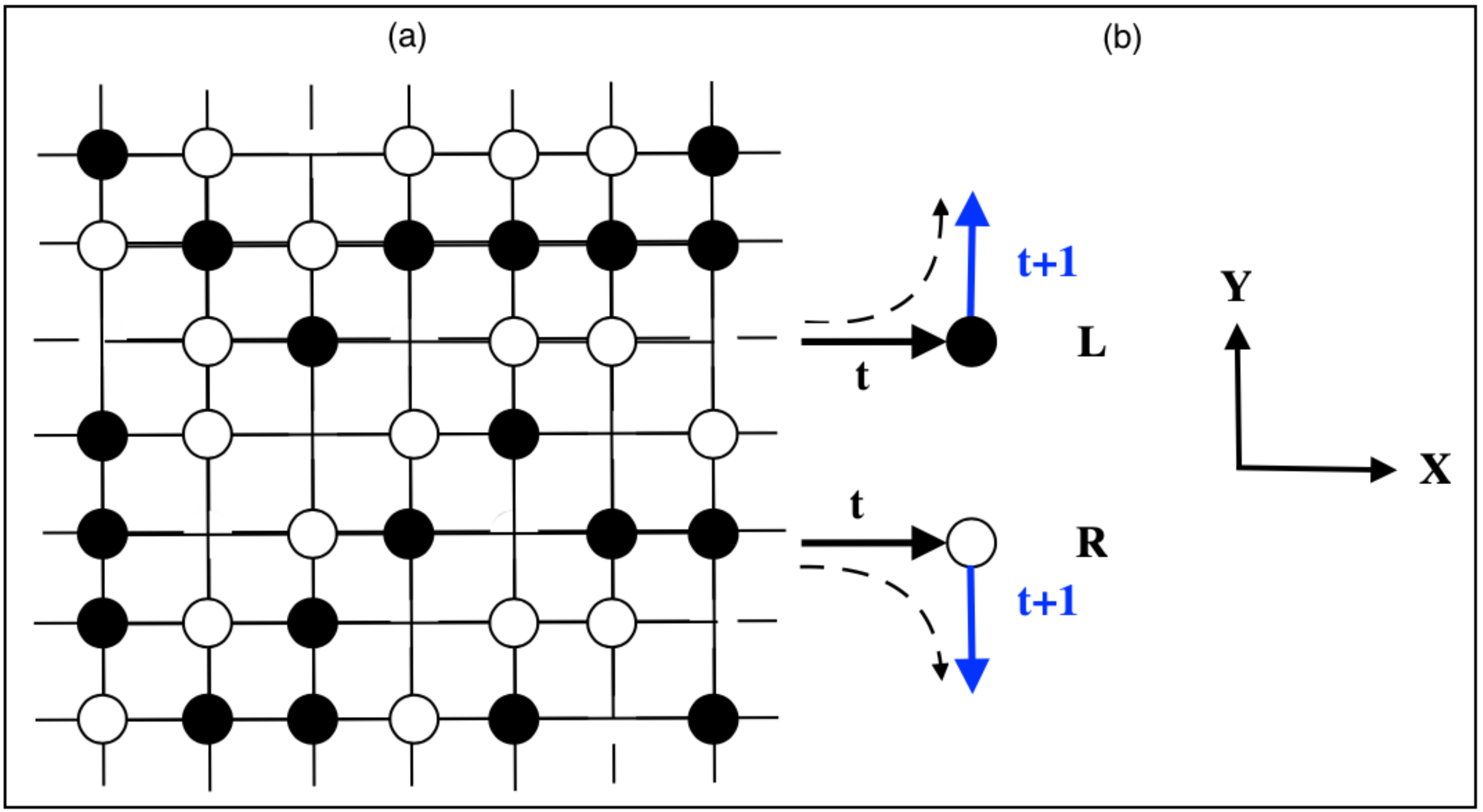}
\caption{Schematic diagram of two dimensional Lorentz Lattice Gas : (a) Initial configuration of LLG with left-rotators (filled circles), right-rotators (open circles) and vacancies (empty sites). (b) State of the particle before ($t$) and after ($t+1$) interaction with rotator; the arrow shows the direction of particle's velocity.}
\label{fig: 1}
\end{figure}

The position of the particle at any time $t$ is defined as ${\bf r}(t) = x(t)\uvec{x} + y(t)\uvec{y}$, 
where $x(t)$ and $y(t)$ are the particle's position (coordinate) along $x-axis$ and $y-axis$ respectively  
(two directions as shown in Fig. \ref{fig: 1}). 
The position of the particle is updated at each time step as, 
$x(t+1)=x(t)+v_{x}(t)$  and  $y(t+1)=y(t)+v_{y}(t)$; 
here $v_i(t)$ $(i=x,y)$ is the $i^{th}$ component of the particle velocity 
at any time $t$ which is updated (when the particle interacts with a rotator) according to the following equation:

\begin{equation}
\centerline{$\left[ \begin{array}{c} v_{x}\left( t +1\right) \\ v_{y}\left( t +1\right) \end{array} \right]\; =\; \left[ \begin{array}{cc} \cos \theta _{s} & \sin \theta _{s} \\ -\sin \theta _{s} & \cos \theta _{s} \end{array} \right]\; \left[ \begin{array}{c} v_{x}\left( t\;  \right) \\ v_{y}\left( t\;  \right) \end{array} \right]$}
\label{eq. 1}
\end{equation}

where the value of $\theta _{s} $ changes according to the type of the rotator that the particle encounters (for $R-$rotator, $\theta _{s} = \pi/2$; for $L-$rotator, $\theta _{s} = -\pi/2$; and for  a vacancy, $V$, $\theta _{s} = 0$).\\

The following quantities are calculated to characterise the properties of the motion of the particle, 
\newline
(i) Mean squared displacement (MSD), which is a measure of {the magnitude of displacement of the particle} from the origin during its motion. MSD ($\Delta(t)$) is defined as 
$\Delta(t)=\langle [{\bf r}(t)-{\bf r}(0)]^2 \rangle$ where $\langle.....\rangle$ \textcolor{black}{represents the averaging over many random, independently generated lattice configurations.}\\
(ii) $P_o(\rho_V,t)$: probability of the particle being in an \textcolor{black}{{\it{open}} (not yet periodic) trajectory} at time $t$, which is the fraction of realizations that are in open trajectories at time $t$ calculated over a large number of realisations. The probability of being in a closed trajectory is denoted as $P_{c}(\rho_V,t)$, and is calculated similarly.

To characterise the motion, we extract the exponent $\beta$ from $\Delta (t)$ by assuming $\Delta(t) \sim t^{\beta}$ and plotting $\Delta(t)$ vs. $t$ on a log-log scale and extracting the slope of the straight line for late times. 
If $\beta>1$, the motion is ballistic. If $\beta \simeq 1$ the motion is diffusive, and if $\beta < 1$, the motion of the particle is subdiffusive. 
The system is studied  for different values of $\rho_{V}$ $(0-0.4)$, upto $10^{5}$ time steps, and over 10000 realizations for each value of $\rho_{V}$. 

\section{Deterministic nature and  closed trajectories}
\label{det}
\begin{figure*}
\centering
\begin{tabular}{cc}
  \includegraphics[width=7cm, height=5cm]{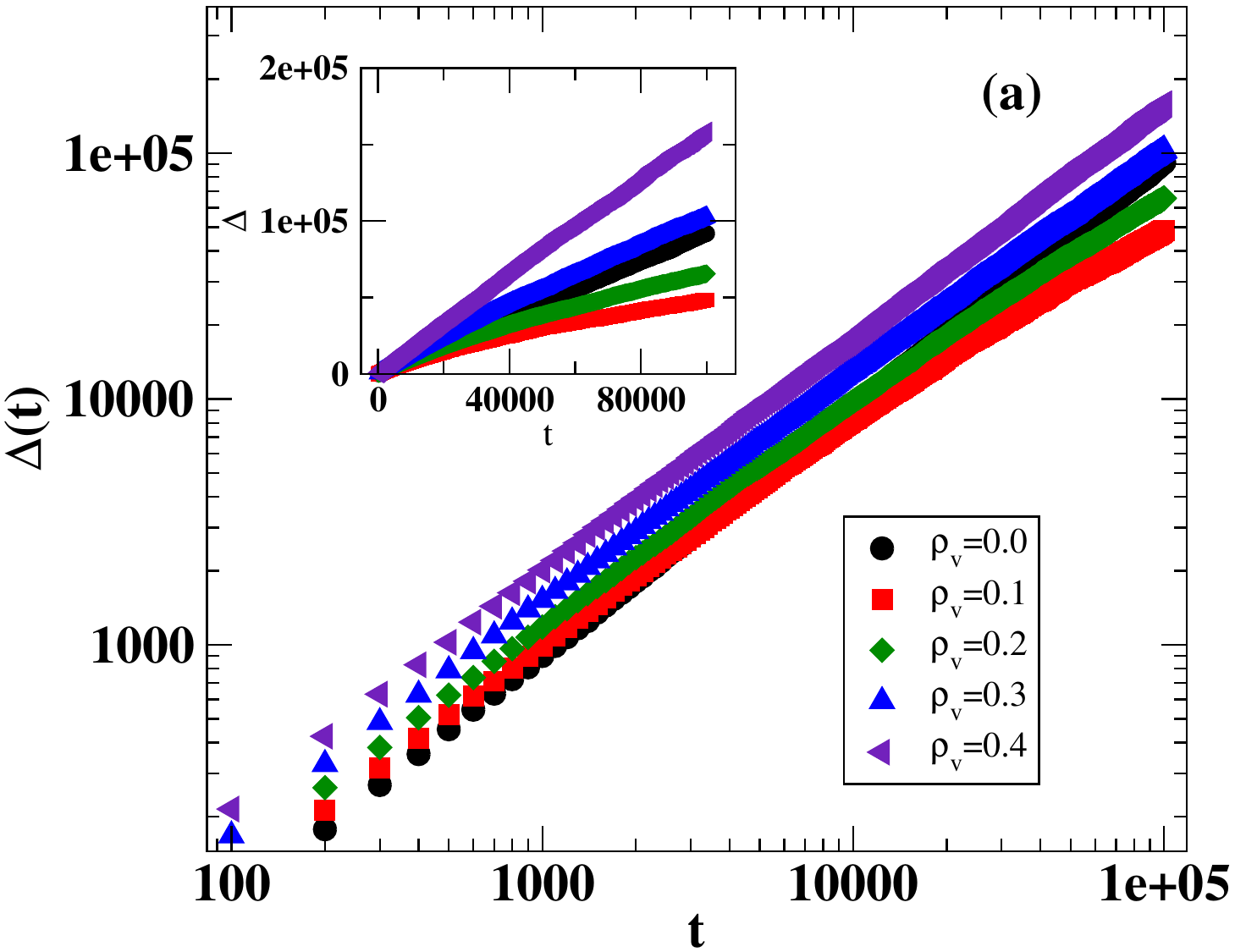} &\includegraphics[width=7cm, height=5cm]{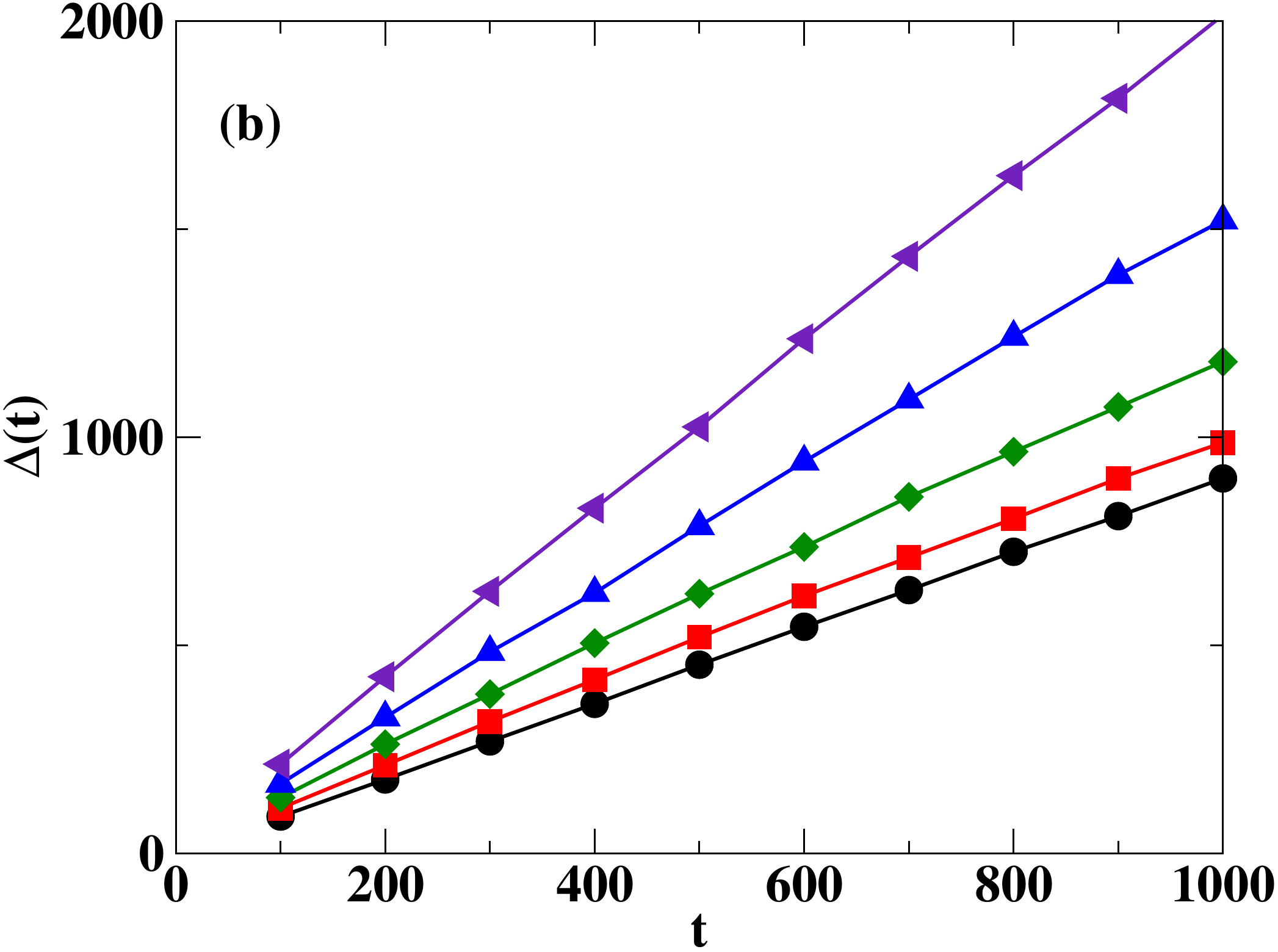}\\
  \includegraphics[width=6.5cm, height=5cm]{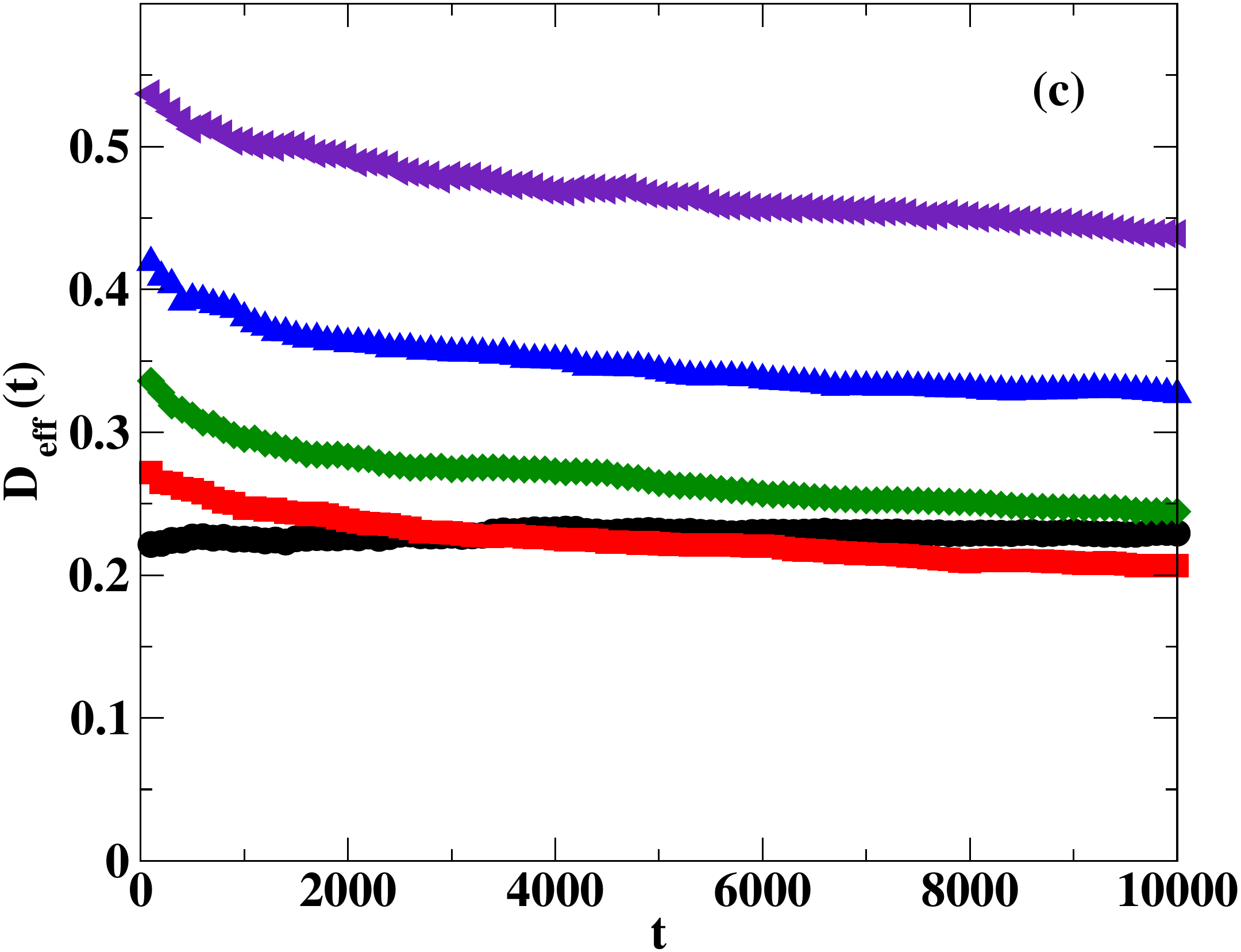} &   \includegraphics[width=6.7cm, height=5cm]{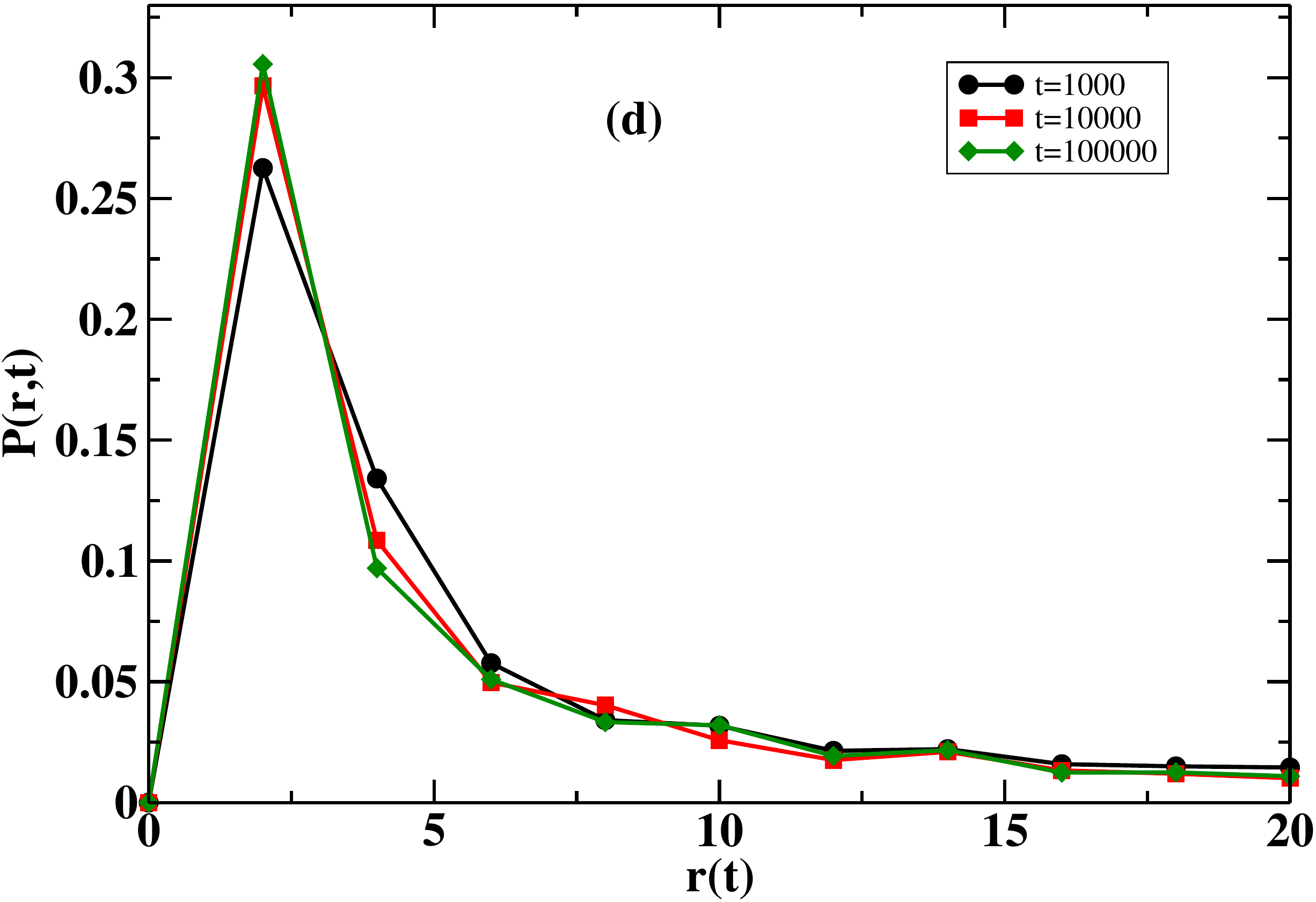} \\
\end{tabular}
\caption{(a) Variation of MSD $\Delta(t)$ with time $(t)$ for different vacancy densities $\rho_V$ (Main). Inset plot shows the crossover from diffusion to subdiffusion. (b) Zoomed in image of plot (a) for $t<1000$. (c) Effective diffusion coefficient $D_{eff}$ vs $t$ for different values of $\rho_V$. (d) $P(r,t)$ vs $r(t)$ at different times for a fully occupied lattice.}
\label{fig: 2}
\end{figure*}
The model we defined in section \ref{model} is  deterministic, and the behaviour of each 
lattice site will be consistent throughout the motion of the particle for a given 
configuration of ``L", ``R" and ``V", regardless of how many times the particle visits a lattice site. 
For example, if the lattice site is occupied by a right rotator for a given configuration, 
the particle will always turn right when it visits that lattice site. However, the results are calculated by averaging the observations across multiple realizations and the random generation of the lattice configuration for different realizations makes the model {\it partially stochastic}.
Hence the particle  {\it deterministically} moves in a {\it random} medium.\\
\newline
The motion of the particle will not be any different from a  random walk on a two-dimensional square lattice until it comes back to a lattice site it has previously visited, as explained in \cite{gunn}. 
When it revisits a site, the particle has to repeat the same turn it had taken when it 
had visited the site earlier in its motion. At this point, some memory is introduced in the motion of the particle. 
The particle will get trapped in a closed trajectory if it revisits a site
with the same velocity. 
 This is because the particle would then leave the site with the same direction of velocity as it did 
in the previous case, and thus it will keep repeating its trajectory, starting from the first point it had revisited. The first such revisited point would have to be the origin. For all values of $\rho_{V}$ (not equal to 1), the  deterministic nature of the model  
lets the particle get stuck in a closed or periodic trajectory aymptotically.\\

For a given value of $\rho_{V}$, the probability of the particle being in a closed trajectory at time $t$ is given as $P_c(\rho_{V},t)$, 
which is calculated by counting the fraction of closed trajectories at time $t$ for a chosen number of realisations. Similarly, the probability of the particle not having been trapped 
in a closed trajectory at time $t$ is given as $P_o(\rho_{V}, t)$. Naturally, $P_o(\rho_{V}, t)+P_c(\rho_{V}, t)=1$.
Although the trajectory of the particle asymptotically becomes a closed trajectory for all values in the chosen range of vacancy density, the aim of our present study is to understand the transient dynamics of the particle and how the density of vacancies changes the approach to such an asymptotic state. Hence our result is divided in two subsections, (i) early time dynamics of particle (first phase of motion) and (ii) late time dynamics (second phase of motion). 
\raggedbottom
\section{Results}
\label{results}

We first investigate the early time dynamics of particle.
Before going to the numerical results, here we briefly
describe the motion in a simple manner. 
When the particle is moving on a lattice with vacancy density $\rho_V$, 
the effective path length ($l_{eff})$ for different values of $\rho_{V}$ is given as $l_\mathrm{eff} = \frac{1}{\sqrt{1-\rho_V}}$. This is greater than or equal to the lattice spacing for a fully occupied lattice (1) if $\rho_V>0$, and increases with an increase in the value of $\rho_{V}$; $l_{eff} \simeq 1.054, 1.118, 1.19, 1.29$ for $\rho_V=0.1, 0.2, 0.3$ and $0.4$ respectively. The probability of a particle encountering a scatterer is effectively $1-\rho_V$, and reduces for a system with a higher density of vacancies, whereas in a fully occupied lattice, the particle always moves through a homogeneously spaced deterministic arrangement of scatterers which it encounters at each timestep. While the reduced frequency of scattering and the larger effective path length before scattering appears to accelerate the motion of the particle when comparing two lattices with non-zero density of vacancies, the same is not true when comparing a fully occupied lattice.
The consistent spacing and frequency of scattering allows a particle in a fully occupied lattice to maintain uniform dynamics throughout its motion. 
Now we characterise the dynamics of particle by measuring the mean square displacement.

\subsection{Mean Square Displacement}

Since all the trajectories close asymptotically, at any given time for a given set of parameters the dynamics of particle can have different characteristics obtained from (i) realizations with closed trajectories, (ii) realizations with open trajectories and (iii) all the realizations. The mean-square displacement of the particle can be broken down into the following functional
form:
\begin{equation}
\label{msd-equation}
{\Delta(t)=\Delta_{o}(t)P_{o}(\rho_V,t)+\Delta_{c}(t)P_{c}(\rho_V,t)}
\end{equation}
Where $\Delta_{o}(t)$ is the mean-square displacement of realizations in open trajectories, and $\Delta_{c}(t)$ is the mean-square displacement of realizations in closed trajectories, both at time $t$. We first study the dynamics of the particle for different values of $\rho_V$ and across all realizations (open as well as closed).\\

In fig. \ref{fig: 2}(a) and (b), we plot $\Delta(t)$ {\em vs}. time for different values of $\rho_V$. The plot of $\Delta(t)$ {\em vs.} $t$ on the $\log-\log$ scale is shown in Fig. \ref{fig: 2}(a)(main), which shows the linear variation of $\Delta(t)$ ($\beta \simeq 1$) with time  $t \simeq 100 - 10000$ for all $\rho_V$. We also observe that, $\Delta(t)$ of a particle increases monotonically with an increase in the value of $\rho_{V}$.\\

In Fig. \ref{fig: 2}(c) we plot the effective diffusion coefficient $D_\mathrm{eff}(\rho_V, t) = \frac{\Delta(t)}{4 t}$ for different values of $\rho_V$. For a particle in a fully occupied lattice, $D_\mathrm{eff} \simeq 0.25$; which is the same value as a particle performing a two-dimensional random walk{\cite{205llg}}. However, due to the deterministic nature of motion, trajectories close with time and the diffusive motion of the particle is not conventional. The radial distance of the particle from the origin at any given time $t$ is denoted as $r(t)$. The probability distribution of radial position $P(r,t)$, does not follow a Gaussian  distribution. The motion is therefore referred to as {\em anomalous} diffusion. The plot of $P(r,t)$ {\em vs.} $r(t)$ for $\rho_V=0$ is shown in Fig. \ref{fig: 2}(d) at three times $t=10^3, 10^4$ and $10^5$. At all times, $P(r,t)$ is a non-Gaussian distribution and the peak of the probability distribution appears close to the origin ($r \simeq 0$).\\

The dynamics of a particle in a lattice with a non-zero vacancy density ($\rho_V \ne 0$) is different from a fully occupied lattice after a large number of time steps. As shown in Fig. \ref{fig: 2}(a)(inset), we observe that $\Delta(t)$ of a particle traversing a partially vacant lattice starts slowing down after a large number of time-steps, and begins subdiffusive motion. It is also observed that a particle traversing a lattice with a lower, but non-zero, value of $\rho_{V}$ slows down much earlier in its motion than a particle in a lattice with a larger value of $\rho_{V}$. We understand this by studying the characteristics of the particle in closed and open trajectories.
\begin{figure*}
\centering
\begin{tabular}{cc}
  \includegraphics[height=5cm, width=6cm]{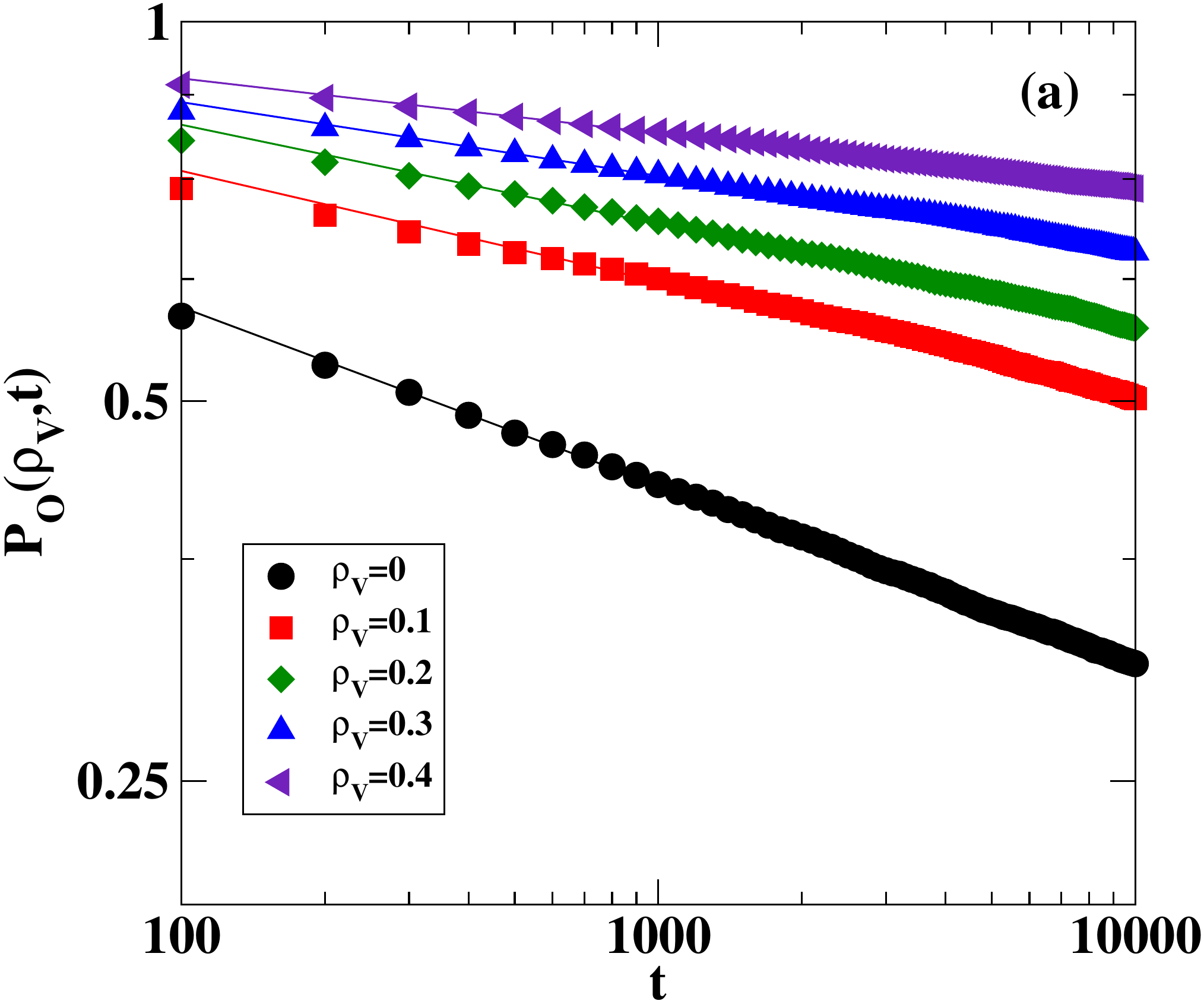} & \includegraphics[height=5cm, width=6cm]{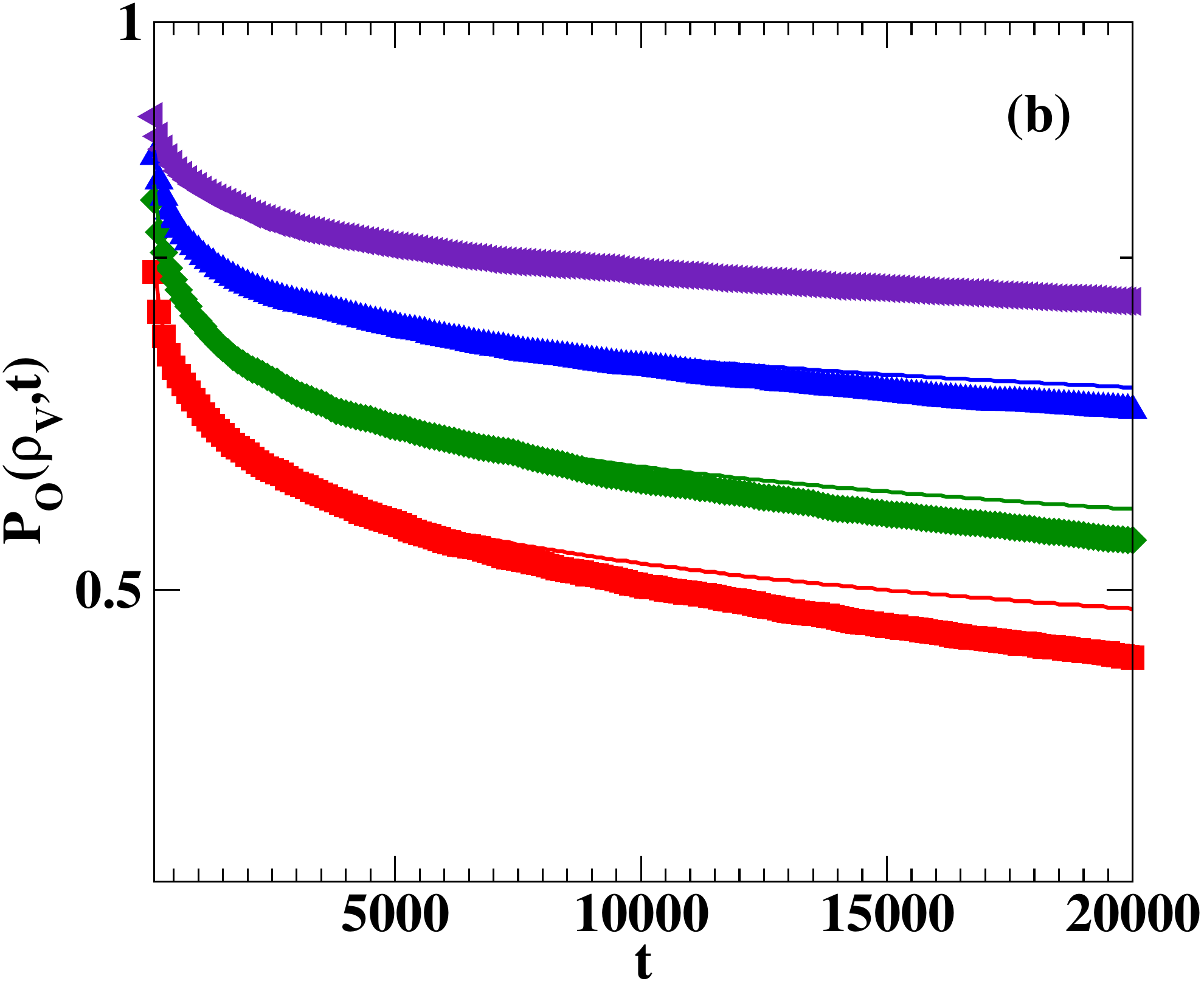}
\end{tabular}
\caption{(a)Variation of probability of being in an open trajectory $P_{o}(\rho_{V},t)$ with time ($t$) on a log-log plot for different values of $\rho_V$ in the first phase, when $P_{o}(\rho_V,t)$ varies as a power law. The straight lines represent the fitted power law functions for each value of $\rho_V$. (b) Variation of $P_{o}(\rho_V,t)$ with time on a semi-log(y) plot during the second phase of motion, when $P_{o}(\rho_V,t)$ decays exponentially with time. As the particles transition from the first phase of motion to the second, $P_{o}(\rho_V,t)$ decays exponentially with $t$ and deviates from the fitted function. }
\label{fig: 3}
\end{figure*}
\subsection{Open and closed trajectories}
The subdiffusive behaviour of $\Delta(t)$ of particles in partially vacant lattices at late times appears due to the closing 
of a larger fraction of trajectories as time progresses.  
We plot the probability of being in an open trajectory $P_o(\rho_V, t)$ as a function of time, for different values of $\rho_V$, as shown in Fig. \ref{fig: 3}(a). We observe that for  time $t\sim 10000$, the probability of being in an open trajectory decays as a power law for all non-zero values of $\rho_V$ (not equal to 1). This period is referred to as the first phase of the particle's motion. In this period, at any point $t$, the value of $P_o(\rho_V,t)$ is higher for higher values of $\rho_V$.\\

The first phase of motion for the particle lasts for a higher number of time-steps when traversing a partially vacant lattice with a higher value of $\rho_V$, but eventually the particle transitions to a second phase of motion, in which $P_o(\rho_V,t)$ decays exponentially as shown in Fig. \ref{fig: 3}(b).\\

We use the form $P_{o}(\rho_V, t) \sim t^{-\gamma(\rho_V)}$, and extract the decay exponent   $\gamma(\rho_V)$ for different values of $\rho_V$ for the first phase of the particle's motion. The value of $\gamma(\rho_V)$ decreases with increasing $\rho_V$. The plot of $\gamma(\rho_V)$ for different vacancy densities  $\rho_V$ in the range $(0, 0.4)$ is shown in Fig. \ref{fig: 4}. It is observed  that $\gamma(0) \simeq 1/7$ and then shows a jump to the value $\simeq 0.1$ and remains  flat up to $\rho_V\leqslant 0.1$, and then decreases sharply.\\
\begin{figure}[H]
\centering
  \includegraphics[width=5.5cm, height=4cm]{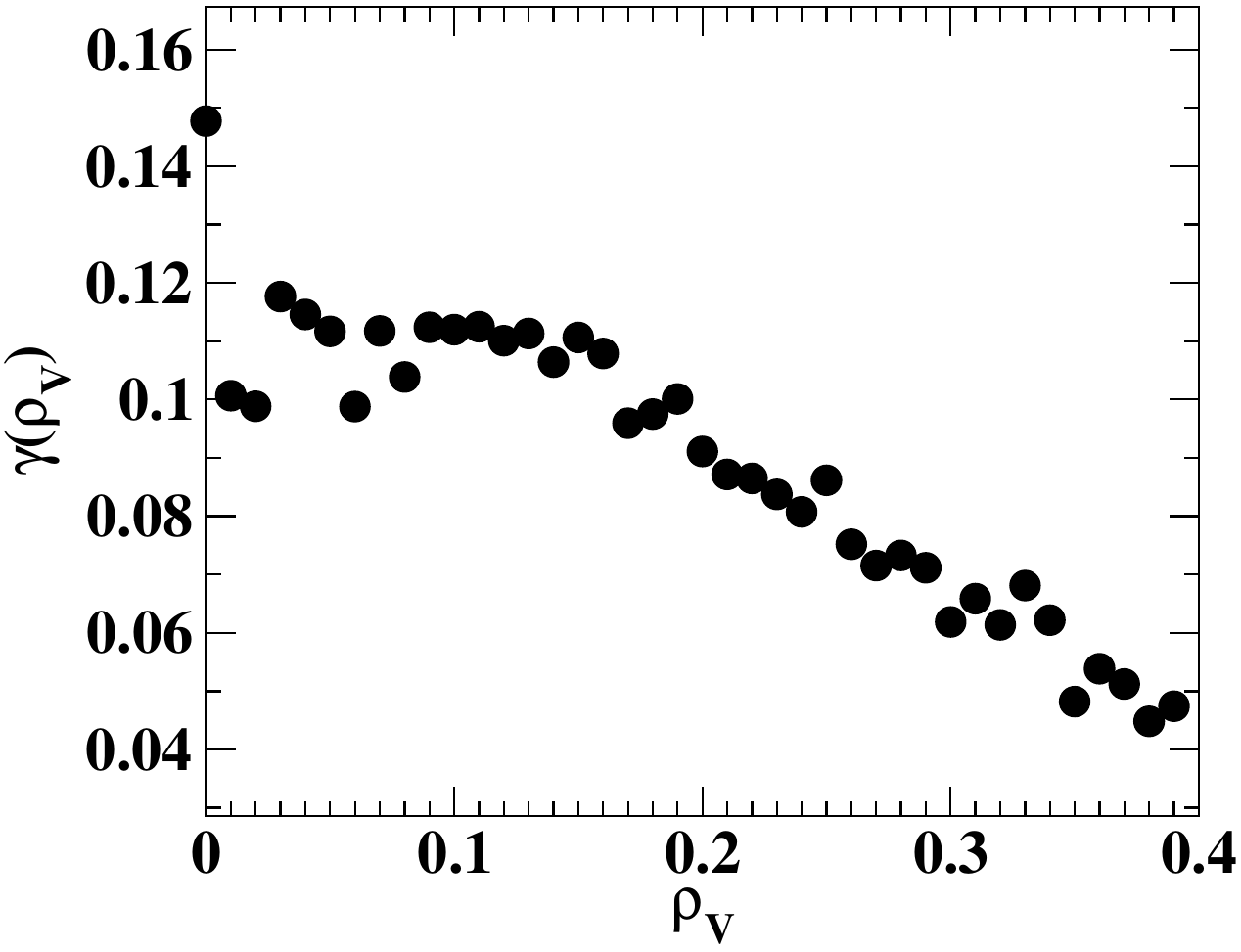} 
\caption{Exponent $\gamma$ vs  $\rho_V$ plot. Value of  $\gamma$ is extracted by fitting power law to $P_0(\rho_V,t)$ vs $t$ plot  in diffusive regime i.e. simulation time  $\mathcal{O}(10^4)$.}
\label{fig: 4}
\end{figure}

\subsection{Properties of Open Trajectories}
\begin{figure*}
\includegraphics[width=17cm, height=5cm]{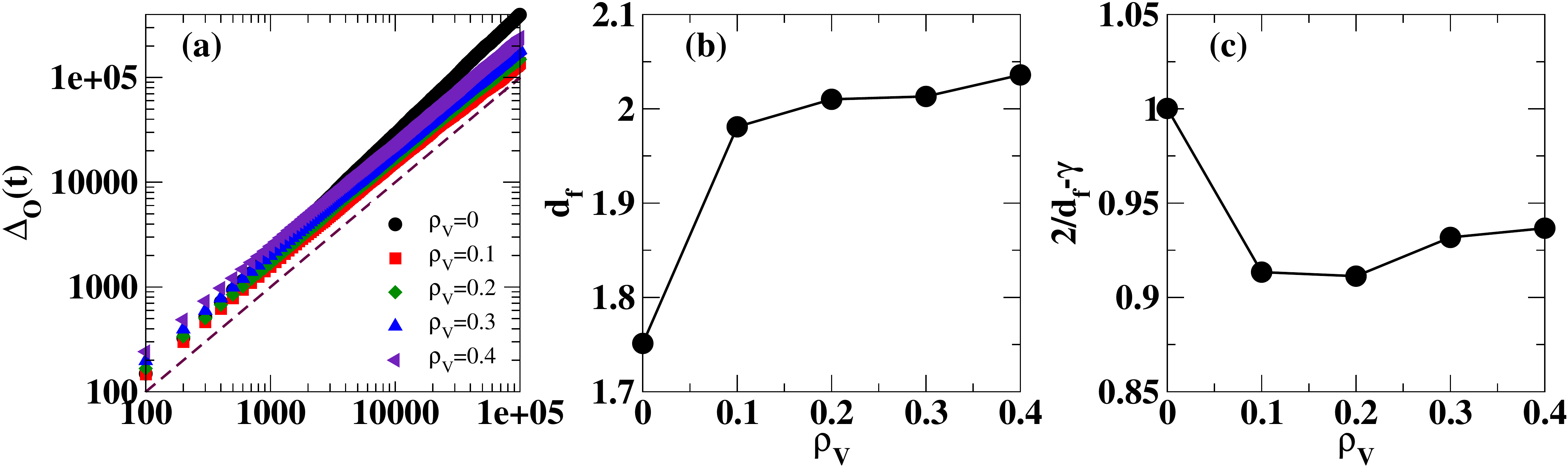}
\caption{Variation of $\Delta_{o}(t)$ with time for different values of $\rho_V$. The dashed line represents a straight line with slope $1$. (b) Fractal dimension and (c) hyperscaling condition ($2/d_{f}-\gamma$) of trajectories for different values of vacancy density in the first phase of motion.}
\label{fig: 5}
\end{figure*}

In Fig. \ref{fig: 5}(a) we plot the mean-square displacement of particle in open trajectories($\Delta_{o}(t)$) as a function of time $t$. The mean-square displacement in open trajectories is larger in a fully occupied lattice than a partially vacant lattice.
We also characterise the nature of the boundary of the trajectory of the particle. The fractal dimension $d_f$ of the particle's motion in the trajectory can be calculated using the asymptotic relation between the distance between two points on a trajectory $R(s)$ and the number of steps separating them $(s)$, \textcolor{black}{which is given as}: 

\begin{equation}
\label{fracd_gen}
{< R^{2}(s)> \sim s^{2/d_{f}}}
\end{equation}

By fixing one of the points to be the origin, the mean-square displacement of the particle on sufficiently large open trajectories (greater than 1000 steps) in the first phase of motion was used to characterise the fractal dimension of the trajectories as follows:

\begin{equation}
{\Delta_{o}(t) \sim t^{2/d_{f}}}
\end{equation}

The variation of $d_f$ for different values of $\rho_{V}$ is given in Fig. \ref{fig: 5}(b). Its value is $7/4$ for motion in a fully occupied lattice, similar to that of a percolating cluster in two-dimensions, and jumps to $2$ for lattices with vacancies. The fractal dimension increases due to the increased crossing of trajectories for increasing vacancy density.\\

The mean-square displacement of closed trajectories after a large number of time steps is much
smaller than the number of time steps. Thus the second term on the RHS in equation (\ref{msd-equation}) has a negligible contribution to the mean-square displacement of the
particle after a large number of time-steps. Thus, the mean-square displacement of the particle
eventually becomes a product of the probability of being in an
open trajectory, and the mean-square displacement of particle in open trajectories. In the first phase of the particle's motion, both these terms vary as a power law with $t$, and therefore, their product also varies as a power-law with $t$ \textcolor{black}{as follows}:\\
\begin{equation}
\Delta_{o}(t)P_{o}(\rho_V,t) \sim t^{2/d_{f}-\gamma}
\end{equation}\\
For a fully occupied lattice, $\Delta_{o}(t)$ is superdiffusive and $\Delta_{o}(t) \sim t^{\alpha}$ with $\alpha \simeq 1.14$, and for particles in lattices with vacancies,  $\Delta_{o}(t) \sim t$.
However, the probability of being in an open trajectory is also lower for a particle in a fully occupied lattice compared to a particle in a partially vacant lattice which is in the first phase of its motion for the same number of time steps. The product of the two factors leads to diffusive motion in the first phase of the particle's motion.\\

Further, the motion of a particle travelling in a fully occupied lattice can be mapped to the bond-percolation problem in two-dimensions \cite{rziff, 205llg}. While the motion in a partially occupied lattice cannot be mapped physically to bond-percolation because of the possibility of the particle crossing over its own trajectory, the power-law exponent associated with $P_{o}(\rho_V,t)$ in the first phase of motion is numerically analogous to the value $(\tau-2)$, where $\tau$ is the critical exponent in the bond percolation problem, such that:
\begin{equation}
P_{o}(\rho_V,t) \sim t^{-\gamma} \equiv t^{2-\tau} 
\end{equation}

In a system of bond percolation \cite{205llg}, 
the critical exponent $\tau$ and fractal 
dimension $d_f$ are related by a hyper scaling relation, which confirms that 
the trajectories are closed asymptotically. The hyperscaling relation is 
\begin{equation}
(\tau-1)d_{f}=2
\end{equation}\\
Using the relation $\gamma \equiv \tau-2$, the condition can be rewritten for this model as:
\begin{equation}
2/d_{f}-\gamma=1
\end{equation}

The expression on the LHS is the power law exponent for the contribution of
open trajectories to the mean-square displacement of the particle. The physical significance of this
condition being satisfied in our model would be the presence of {\it diffusive} 
motion. As seen in Fig. \ref{fig: 5}(c), a fully
occupied lattice exactly satisfies this condition and as a result, shows anomalously diffusive
behaviour throughout its motion. However, we see a  deviation from this condition for
systems having a partially vacant lattice. For the values of $\rho_{V}$ considered in this paper, the deviation is \textcolor{black}{maximal} near $\rho_V=0.1$ and then 
reduces as we increase the density of vacancies.
The deviation from the exact value, even before the exponential decay of $P_{o}(\rho_V,t)$, is also responsible for the slowing of the growth of $\Delta(t)$ (as shown in Fig. \ref{fig: 2}(a)) in partially vacant lattices, after a large number of time steps.\\

\section{Discussion}
\label{discussion}

We studied the dynamics of a single particle moving on a two-dimensional Lorentz-lattice gas. The lattice is occupied by right/left rotators and some of the sites are left vacant. The particle moves along the bond of the lattice and turns left/right if it encounters a left/right rotator, or continues its motion if it passes through a vacancy. The dynamics of particle motion is explored for various vacancy densities. Although the probability of the particle being in an open trajectory, $P_o(\rho_V,t)$, eventually decays to zero if the system is allowed to run for a large number of time steps ($t \rightarrow \infty$) for all values of $\rho_V$ (not equal to 1), the transient state dynamics are different for a fully occupied lattice ($\rho_V=0$), and lattices with a non-zero vacancy density. For a fully occupied lattice, the dynamics are anomalously diffusive throughout its motion; the mean square displacement varies linearly with time, but the probability distribution of the particle's radial distance is non-Gaussian. The probability of being in an open trajectory at time $t$ also decays as a power law throughout the motion of the particle. However, for any lattice with a non-zero vacancy density, the motion of the particle is described by two phases. In the first phase of its motion, the motion of the particle is similar to the motion of a particle in a fully occupied lattice; the mean square displacement grows linearly with time and the probability of the particle being in an open trajectory decays as a power law with time. The particle considerably slows down in the second phase of its motion. 

The second phase of the particle's motion is characterised by an exponential decay of $P_{o}(\rho_V,t)$ \cite{205llg}, which now vanishes at a much faster rate than the first phase. The mean-square displacement of realizations with particles in open trajectories continues to grow linearly with time, and the product of the two factors will eventually become a monotonically decreasing function. As $P_{o}(\rho_V,t) \rightarrow 0$, the motion of the particle is only characterised by oscillating, periodic closed trajectories. Thus, the value of the $\Delta(t)$ for particles in partially vacant lattices will eventually oscillate about a constant value.

Further, the boundary of trajectory of the particle, displays a fractal motion with fractal dimension $\sim 7/4$ when traversing a fully occupied lattice. However, the value of its fractal dimension shows a jump when it goes to a partially vacant lattice and asymptotically approaches the value $2$.
 
In the current study, we focused on the dynamics of a single particle. A system involving multiple particles, with interactions between the particles, may dramatically change the results of the study. Also, our present study is limited to isotropic bond lengths, it would be interesting to study the dynamics on an anisotropic model \cite{gunn}

\section{Acknowledgement}
SM would like to thank E. G. D. Cohen  for introducing the problem of the dynamics of particle on Lorentz lattice gas and S. S. Manna for useful discussion. SM and SK also thank DST,  SERB(INDIA),  project  no.ECR/2017/000659 for partial financial support. 

\end{document}